\documentclass[useAMS,usenatbib]{mn2e}
\usepackage{graphicx}

\title[An analysis of HCN observations of the Circumnuclear Disk at the galactic centre]{An analysis of HCN observations of the Circumnuclear Disk at the galactic centre}

\author[I.L.Smith and M.Wardle]
  {I.L.Smith$^1$\thanks{E-mail:ian.smith@mq.edu.au}
  and M.Wardle$^1$ \\
$^1$Department of Physics and Astronomy \& Research Centre for Astronomy Astrophysics and Astrophotonics ,\\ Macquarie University, Sydney NSW 2109 Australia} 
\begin{document}

\date{ }

\pagerange{\pageref{firstpage}--\pageref{lastpage}} \pubyear{2012}

\maketitle

\label{firstpage}

\begin{abstract}
\paragraph*{}

The Circumnuclear Disk (CND) is a torus of dust and molecular gas rotating about the galactic centre and extending from approximately 1.6pc to 7pc from the central massive black hole, SgrA*. Large Velocity Gradient modelling of the intensities of the HCN 1-0, 3-2 and 4-3  transitions  is used to infer hydrogen density and  HCN optical depth. From HCN observations we find the molecular hydrogen density ranges from 0.1 to 2 $\times$ 10$^{6}$ cm$^{-3}$, about an order of magnitude less than inferred previously. The 1-0 line is weakly inverted with line-centre optical depth approx $-$0.1, in stark contrast to earlier estimates of 4. The estimated mass of the ring is approximately 3 $-$ 4 $\times$ 10$^{5}$M$_{\odot}$ consistent with estimates based on thermal dust emission. The tidal shear in the disk implies that star formation is not expected to occur without some significant triggering event. 

\end{abstract}


\section{Introduction}
\paragraph*{}
The Circumnuclear Disk (CND) is a ring of gas and dust located close to and around the Milky Way's galactic centre.  The CND was first discovered in IR continuum emission from dust at 30, 50 and 100$\mu$m using the Kuiper airborne observatory \citep{Becklin1982}. The CND appeared as two lobes that were symmetrically located to the NE and SW about SgrA$^{*}$ showing as 30$\mu$m emission close to the galactic centre in a relatively dust-free central cavity. The symmetry and orientation of the lobes suggested a ringlike structure with a major axis approximately aligned with the galactic plane. Observations of CO, CS and HCN  subsequently led to the discovery of the CND rotating about the galactic centre \citep{Serabyn1985,SGW1986,Guesten1987}. The disk was found to have an inner radius of 1.5 to 1.7pc and extend to 5pc in HCN and $>$7pc in CO \citep{Guesten1987}; more recent observations have detected HCN out to 7pc and CO up to 9pc from the galactic centre. \citep{Chris2005,Oka2007,Oka2011}.

\paragraph*{}
The ring material orbits in the combined gravitational potential of the central stellar cluster and the approx 4$\times$10$^{6}$ M$_{\odot}$ black hole SgrA$^{*}$ with has a rotational velocity of $\sim$110 km s$^{-1}$ between 2 to 5pc from the galactic centre \citep{Ghez2005,Genzel2010}.
%(see Fig.\ \ref{Velmap}).
Lower velocities are indicated by CII and CO(7-6) at radii $\geq$ 4pc and higher velocities, 130--140 km s$^{-1}$, are indicated by HCN in the North Eastern part of the ring at 2pc \citep{Guesten1987}. \citet{Marshall1995} fitted a 3D rotating ring model to HCN (4-3) and (3-2) data and inferred a flat velocity profile, while noting that \citet{Harris1985} showed a velocity fall off between 2 and 6 pc from the centre in the CO 7-6 line.

\paragraph*{}
The disk is clumpy, with clump diameters varying from 0.14 to 0.43 pc and gas and dust rotating in a number of kinematically distinct streams about the galactic centre \citep{Guesten1987,Jacks1993}. The disk's major axis is aligned to a position angle of $\sim$ 25$^{\circ}$ and inclination of $\sim$ 70$^{\circ}$ to the plane of the sky. \citep{SGW1986,Jacks1993,Marshall1995}. It was noted that the rotation was perturbed in several ways with a large local velocity dispersion throughout the disk. The position angle changes with radius and in inclination with azimuthal angle, i.e. the disk is warped. The perturbations together with the disk's clumpiness indicate a non-equilibrium configuration with an age of only a few orbital periods \citep{Guesten1987,Genzel1989}, i.e. $\sim$10$^{5}$ years.

\paragraph*{}
The far infrared (FIR) continuum emission is well represented by thermal emission from dust at 20, 60 and 100\,K. Observations by \citet{Mezger1989} showed that warm dust $\geq$60\,K accounts for only $\sim$10\% of the total mass in the CND and is located near the ionisation front of SgrA West. \citet{Etxa2011} found the spectral energy distribution for the FIR emission from dust in the CND was best represented by a continuum summing the contribution from dust emission at temperatures of 90, 44.5 and 23\,K with the cold component accounting for $\sim$ 3.2$\times$10$^{4}$ M$_{\odot}$ out of the estimated total mass of $\sim$ 5$\times$10$^{4}$ M$_{\odot}$ in the central 2pc of the CND and is similar to the findings of \citet{Mezger1989}. The remaining dust in the disk appears to be rather cold at $\sim$20\,K and similar to dust temperatures in clouds located in the inner 100pc of the Galaxy \citep{P-Price2000}. 

\paragraph*{}
Estimates of molecular hydrogen densities, n(H$_{2}$), for the clumps have progressively increased from 3$\times$10$^{5}$ in the 1980's \citep{Genzel1989} to 10$^{5}$ and 0.4 to 5$\times$10$^{6}$ \citep{Jacks1993,Marr1993} then 3-4$\times$10$^{7}$ cm$^{-3}$ \citep{Chris2005}. However \citep{Genzel2010} has raised questions about the molecular density of the CND's clumps and summarises the position by describing the two prevailing scenarios as
\begin{enumerate}
\item the original view of less dense (10$^{6}$cm$^{-3}$) warm gas ($>$ 100\,K) clumps which are tidally unstable with a transient lifetime of $\sim$10$^{5}$ yr, or  
\item  the more recent idea of denser (10$^{7}$-10$^{8}$ cm$^{-3}$) cool gas (50-100\,K) residing in  stable clumps with long lifetimes $\sim$10$^{7}$yr, that is long enough for the opportunity for star formation from clump condensation. 
\end{enumerate}

\paragraph*{}

Modelling by \citet{Subr2009} of a coherently rotating stellar ring within the CND shows that a CND mass of $\sim$ 10$^{6}$ M$_{\odot}$ would destroy the stellar ring by gravitational torque within its estimated lifetime of 6\,Myr, provided the inclination of the stellar ring varies by more than 5$^{\circ}$ from 90$^{\circ}$ with respect to the inclination of the CND. Infra-red emission from the dust in the CND has the characteristics of an optically thin medium and is further evidence for scenario one.  For an optically thick medium (scenario two) clumps would appear as dark spots in an infra-red image \citep{Mezger1996}. To date there have been no recorded observations of such dark spots. The dense $\>$ 10$^{7}$ cm$^{-3}$ hydrogen scenario two, relies on observations of HCN 1-0 \citep{Chris2005} and HCN 4-3 \citep{MMC2009} which conclude that the average size  of 0.25pc  and hydrogen density $\sim$10$^{8}$\,cm$^{-3}$ of the clumps can provide stability against tidal forces and result in long lifetimes. 

\paragraph*{}
This paper uses a Large Velocity Gradient (LVG) model to analyse the results from papers based on  observations of three HCN transitions, 1-0, 3-2 and 4-3, to support the lower $\sim$10$^{6}$\,cm$^{-3}$, hydrogen density scenario and is arranged as follows :-
\begin{itemize}
\item Section \ref{CND Props} establishes the properties and orientation of the disk and its relation to surrounding features based on publications of numerous authors.
\item Section \ref{clumpselect} describes how two groups of clumps observed in the HCN 1-0,3-2 and 4-3 transitions and the HCO$^{+}$ 1-0 transition are selected for analysis.
\item Section \ref{LVG} describes the LVG model used to analyse the selected clumps.  
\item Section \ref{Results} reports the results of modelling the two groups of clumps.
\item Section \ref{anal} presents an analysis of the results for both groups of clumps that shows an optically thin HCN 1-0 transition and a clump density of about 10$^{6}$cm$^{-3}$ consistent with the first scenario described above.
\item Section \ref{discuss} discusses the flaw in the argument presented by \citet{Marr1993} that molecular gases with the same kinetic temperature and share the same physical space have equal excitation temperatures. This conjecture is then used to argue an optically thick (4) H$^{12}$CN 1-0 transition based on the equality of the [H$^{12}$CN]/[H$^{13}$CN] abundance ratio to the ratio of their respective optical depths.
\item Section \ref{concl} concludes that the CND is an optically thin,
relatively low density gas ring which is likely to have a short lifespan due to
disruption by tidal shear forces generated by the black hole and stellar material inside the inner radius of the ring.
\end{itemize}

\section{CND's Physical Properties and Related Objects} \label{CND Props}

\subsection{HCN(1-0) Clumps}

\paragraph*{}
\citet{Chris2005} list twenty-six HCN (1-0) clumps, labelled A-Z, with their size, central velocity, width and integrated flux. The criteria for a clump's inclusion was that it was a bright emission source which was isolated in position and velocity space. Their list is not exhaustive but is a good representative sample containing the majority of bright sources and a few lower emission sources. The sample was also restricted to clumps in the CND, except for clumps X and Y which are located in the linear filament which is a feature adjacent to the north-western side of the CND.

\paragraph*{}
We use the HCN(1-0) clumps catalogued by \citet{Chris2005} to determine the CND's orientation in the sky by deprojecting their RA and Dec offsets from SgrA$^{*}$ see \ref{cordtran} to deprojected offsets with Eqn. \ref{cordtran}. The deprojected distances from SgrA$^{*}$ are compared with those tabulated by \citet{Chris2005} to confirm that our CND model agreed with \citet{Chris2005}. Figs.\ \ref{projdeproj} (a) \& (b) show the projected and deprojected views of the CND.

\paragraph*{}
A constant disk rotational velocity is used to predict clump line of sight velocities for comparison with observed velocities to determine which clumps lie within the rotating disk.

\paragraph*{}
The predicted line of sight (los) velocities for the observed clumps are based on a constant rotational velocity (110kms$^{-1}$) for the disk and the clumps being located in the mid plane. This model has been adopted by a number of authors including \citet{Marshall1995} and \citet{Guesten1987}. In \citet{Harris1985} a decline in rotational velocity by a factor of 1.4 to 2 is assumed between a disk radius of 2 to 6pc, which is consistent with a ``Keplerian'' (R$^{-\frac{1}{2}}$) decline. Here we adopt the flat velocity model on the basis that the majority of clumps are within 2pc of SgrA$^{*}$ where the rotational velocity is considered constant and that the differences in los velocity produced by a declining rotational velocity would be negligible in any case due to the small variation in distances from  SgrA$^{*}$. Velocities along the disk's radii and normal to the disk's plane are assumed to be zero. Fig \ref{radvmaser} shows the comparison of predicted with observed los velocities.

\paragraph*{}
\citet{Chris2005} list the PA of the CND's projected major axis as 25$^{\circ}$ and the tilt of the disk's axis of rotation to the plane of the sky as varying between 50$^{\circ}$ and 75$^{\circ}$. We found that a PA of 25$^{\circ}$ and tilt angle of 60$^{\circ}$ produced calculated distances that agreed closely with the deprojected clump distances of \citet{Chris2005} and so were adopted for this paper.

\subsection{Water, Methanol and OH Masers} \label{secmasers}
\paragraph*{}
This subsection collates the positions of water (22 GHz) and methanol (44 GHz) masers in and around the CND observed using the Green Bank telescope \citep{FYZ2008} and 
OH (1720 MHz) masers from papers by \citet{Karlsson2003,FYZ1999} based on VLA observations from 1986 to 2005 together with new observations by \citep{Sjman2008}. The maser positions are based on co-ordinates from \citet{Sjman2008} and \citet{FYZ2008} and are shown in Fig.\ \ref{projdeproj}.

\paragraph*{•}
The methanol masers identified by \citet{FYZ2008} near clumps F, G and V are
marked by green squares on Figs.\ \ref{projdeproj} and \ref{radvmaser}. Both the
projected and deprojected plots confirm the masers are in the vicinity of their
respective clumps and their observed los velocities are within 10km s$^{-1}$ of
their respective clumps. The masers near clumps F and G are part of a group of
eight methanol masers that are on the eastern side of the CND, while clump V is
close to the inner western edge of the CND and is a site of shocked H$_{2}$
emission. A red shifted wing of HCN emission from clump V in the direction of
the methanol maser could be the signature of a classic one sided outflow as
occurs in star forming regions \citep{Mehringer1997}. \citet{FYZ2008} propose
that the Class I Methanol masers close to the three HCN clumps are evidence of
protostars about 10$^{4}$ years after gravitational collapse.  
Water masers (red crosses) were found close to clumps B-Z, D, E, F, O and W. 

\paragraph*{}
\citet{Sjman2008} identified OH masers near clumps B and N (cyan diamonds in Figs.\ \ref{projdeproj}(a) and (b) and \ref{radvmaser}). The two masers in the NW lobe have highly positive los velocities of +132km s$^{-1}$ which closely match the observed los velocity of +139kms$^{-1}$ for clump B. The masers near clump N have highly negative los velocities, --141 and --132km s$^{-1}$, compared with --64km s$^{-1}$ for clump N so appear unrelated to this clump while associated with the two masers in the NW lobe. Two other OH masers with los velocities of --104 and --117km s$^{-1}$ lie about a parsec outside clump O which has a los velocity of --108 km s$^{-1}$ and could be in the same rotating stream.  We agree with \citet{Sjman2008} that the high velocities together with their symmetry of positive and negative values indicate that these masers are rotating in the CND structure and that the source of excitation is collisions of CND clumps and is not related to the supernova shell of SgrA East. The clump of OH masers SE  of the CND are an indication of interaction between the expanding supernova shell and the +50km s$^{-1}$ molecular gas cloud.

\subsection{HCN Clump and Maser Positions and Line of Sight Velocities} \label{corevels}
\paragraph*{}
Fig.\ \ref{radvmaser} shows predicted los velocity curves based on combinations of  PA (clockwise from East) 60$^{\circ}$ and 70$^{\circ}$  and inclination angles of the disk's axis of rotation, 60$^{\circ}$ and 75$^{\circ}$ to produce an envelope of curves that constrain the range of predicted los velocity values. The clump positions and observed los velocities are superimposed for comparison with the curves and an assessment made as to the likelihood of particular clumps being located in the disk. 

\begin{figure*}
\begin{minipage}{140mm}
\includegraphics[scale=0.75,bb= 50 50 500 880,trim=30 100 30 200,angle=90] 
%{HCN_Cores_and_Masers_J2000_v2.ps} \fontsize{9} {9} %\caption[HCN Core and Maser Fig 1
{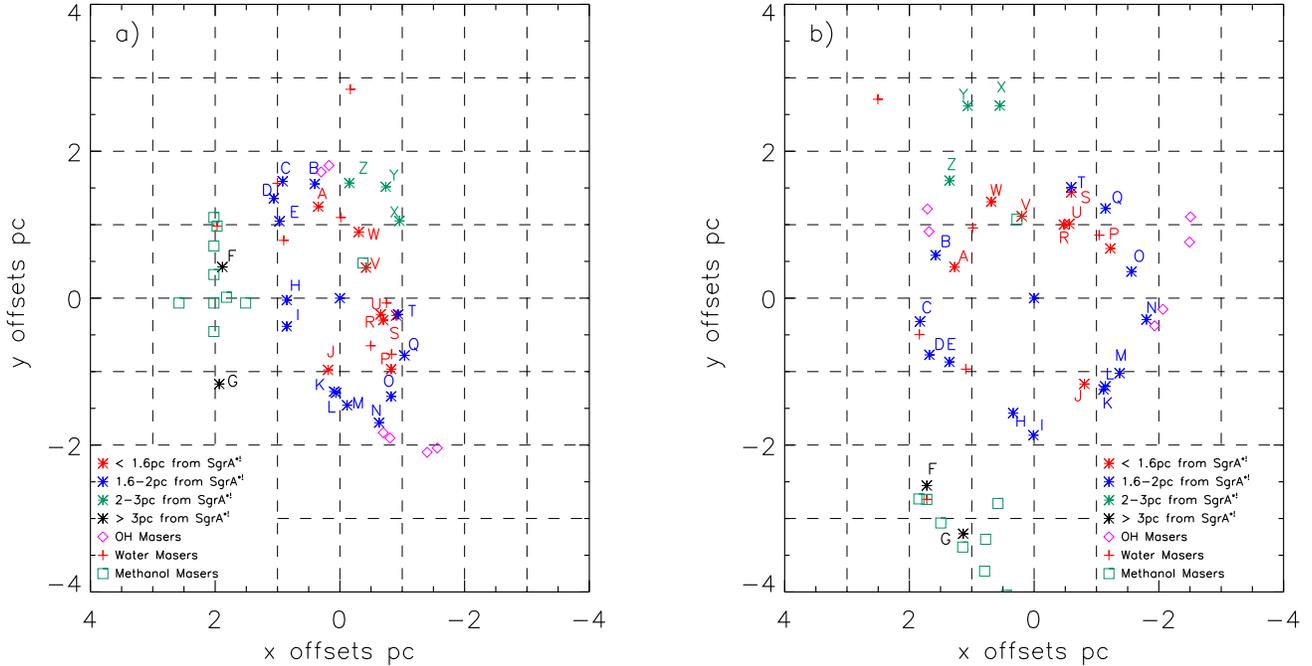} \fontsize{9} {9} 
\caption[HCN clump and Maser Projected and Deprojected Plots ]{\label{projdeproj} a) is the sky view of CND HCN(1-0) clumps (A to Z), water, methanol and OH masers positions relative to the origin SgrA$^{*}$. X and Y axes oriented in direction of increasing RA and Dec (J2000). \\
b) is the deprojected view of HCN(1-0) clumps and masers and lie in the disk's plane. Masers are only plotted  where the deprojected distances from SgrA$^{*}$ are less than 4pc and their observed los velocities are compatible with CND los velocities.}

\end{minipage}
\end{figure*}

\begin{figure*}
\begin{minipage}{160mm}
\includegraphics[scale=0.60,bb= -110 179 700 665,clip=true,trim= 10 0 10 0]
{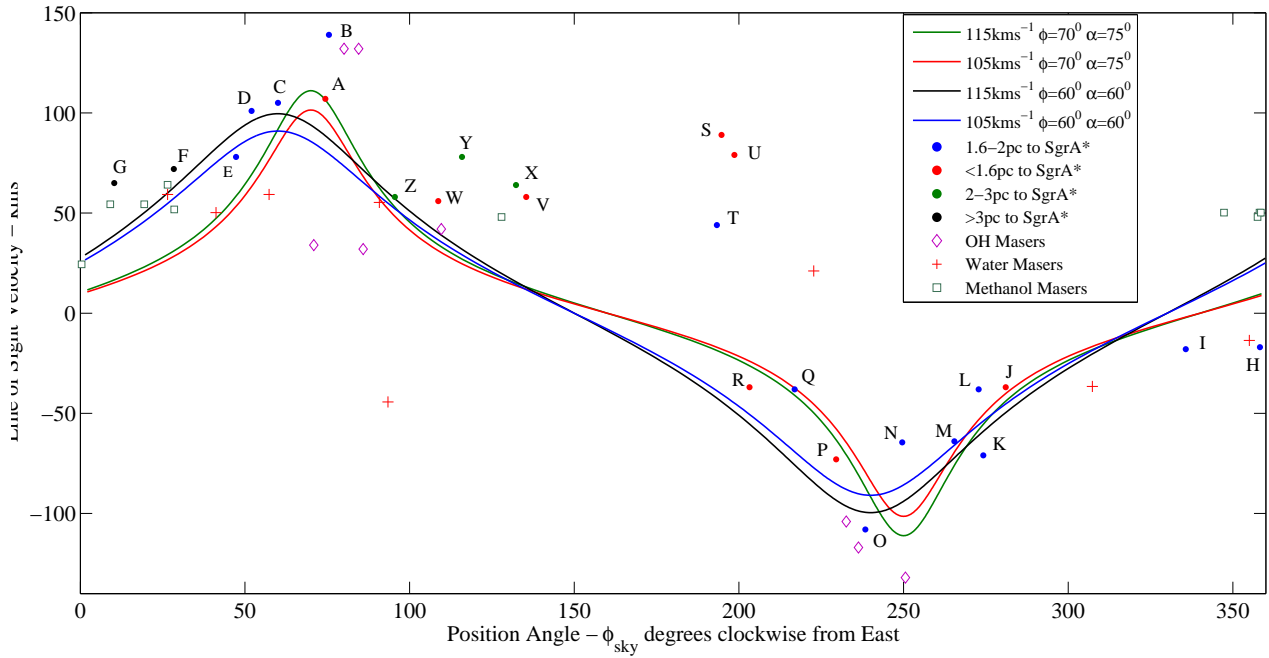} \fontsize{9} {9} 
\caption[Plot of HCN Clump and Maser Radial Velocities]{\label{radvmaser} 
Plots of predicted line of sight velocities generated by varying the PA of the disk's major axis , $\phi$, and its inclination $\alpha$. As listed in the figure's legend, rotational velocities of 105 and 115 km s$^{-1}$ were adopted as the upper and lower limits bracketing 110 km s$^{-1}$ (see Table \ref{Ring pars}) . Increasing  PA shifts the curve horizontally to the left. Increasing the tilt angle increases the peak value of the los velocity. The observed los velocities of HCN(1-0)clumps (A to Z) by \citep{Chris2005}, water, methanol and OH masers are also plotted.} 
\end{minipage}
\end{figure*}

\paragraph*{}
The right hand plot of Fig.\ \ref{projdeproj} indicates that only five clumps (F, G, X, Y and Z) lie outside a distance of 2pc from SgrA$^{*}$ and eight clumps (A, J, P, K, S, U, V and W) lie within 1.6pc from SgrA$^{*}$. Six of the clumps, (A, C, D, E, F and Z), are located in the NE section of the ring and nine clumps, (J, K, L, M, N, O, P, Q and R) are in the SW section. This indicates that most of the detected HCN clumps are located in the inner section of the CND (i.e.\ within 2pc of SgrA$^{*}$). 
 
\paragraph*{}
Fig.\ \ref{radvmaser} shows that the los velocities of eight clumps (B, H, S, T, U, V, X and Y) have discrepancies with model predictions, in excess of 35km s$^{-1}$. All these clumps, except U and V, lie between $\sim$ 1.6 and 3.5pc from SgrA$^{*}$. Five clumps (H, S, T, U and V) are located within a deprojected distance of $\sim$ 1.6pc of SgrA$^{*}$ and may be influenced by the movement of ionised gas in the western arm of the mini-spiral which has positive los velocities at these positions see \citet{Zhao2009} in contrast to the observed mainly  negative velocities of these clumps by \citet{Chris2005}. Clumps X and Y are located in the linear filament that is located adjacent to the NW section of the CND. Clump B is located in the Northern Lobe close to the ring's northern gap. The Northern Arm of the mini spiral is in the vicinity about 0.2pc to the west and 0.3pc to the South. Elements N1 and N2 of this feature have observed los velocities of +78 and +100 km s$^{-1}$ respectively, see Table 3 of \citet{Zhao2009}, compared to the observed +139 km s$^{-1}$ for clump B. Assuming clump B and the two methanol masers are part of the CND requires that they be located in a CND streamer circulating at a much higher rotational velocity ($\sim$ 150 km s$^{-1}$) than the average 110 km s$^{-1}$. clumps F and G are two outlier clumps at deprojected distances more than 3pc east of the galactic centre and some 2pc inside the NE group of methanol  masers reported by \citet{Sjman2008} as marking the shock front of the SgrA East supernova remnant (SNR) shell.

\paragraph*{}
 Following the above considerations clumps D,I,M,O,P,W and Z were then selected for analysis using the LVG model.

\section{Selection of HCN Clumps for Analysis} \label{clumpselect}
\paragraph*{}
LVG analysis is complicated by the lack of data for multi transitions of a trace molecule from a single source. Data for sources that is available has different spatial resolutions since it has been observed with different telescopes at different wavelengths. The analysis relies on modelling emission from three transition levels to produce reliable HCN column densities, hydrogen densities and opacities for the clumps.
A literature review led to a choice of two groups of clumps that had been observed in multiple transitions of HCN and that were physically and kinematically related.

\paragraph*{}
The first group of five clumps were collated and labelled A to E by \citet{Marr1993} who observed clumps in H$^{13}$CN 1-0 and HCO$^{+}$ 1-0 and H$^{12}$CN 1-0 \citep{Guesten1987} channel maps were convolved with a 12$\times$12 Gaussian beam before extracting spectra to allow direct comparison with HCN 3-2 \citep{Jacks1993} spectra. All the data was produced from unresolved images but had a consistent set of related intensities, spatial and kinematic properties which were modelled by \citet{Marr1993}. Our modelling is in effect a re-evaluation of the earlier analysis. 

\paragraph*{}
\citet{Marr1993} indicate the rms uncertainties for H$^{12}$CN,
H$^{13}$CN and HCO$^{+}$ brightness temperatures are 0.2K which represents an uncertainty for peak brightness temperatures ranging from 4 to 20\%. The peak brightness temperatures for HCN 3-2 sourced from \citet{Jacks1993} have a 30\% uncertainty.
\paragraph*{}
The dilution factors for clumps A to E were calculated using their corresponding \citet{Chris2005} clump sizes divided by the 12$\times$12 beam of the telescope used for the HCN 3-2  \citet{Jacks1993} observations. Clump D is not apparent in the 3-2 observations so the average \citet{Chris2005} clump size of 0.25pc was used to calculate its dilution factor. The dilution factors for the clumps were used to convert the observed peak brightness temperatures in Table 1 of \citet{Marr1993} to resolved temperatures in Table \ref{Gp1}.

\begin{table*}
\begin{minipage}{145mm}
%\tiny
\caption{\label{Gp1} Properties of 1st Clump Group}
\begin{tabular}{ccccccccc}
\hline
\multicolumn{2}{c}{Clump}&\multicolumn{1}{c}{Tracer}&\multicolumn{1}{c}{Resolved}&\multicolumn{1}{c}{Central}&\multicolumn{1}{c}{Line Width}&\multicolumn{1}{c}{Dilution}&\multicolumn{2}{c}{Maximum Brightness} \\
\multicolumn{1}{c}{Labels\footnote{clump IDs from \citet{Marr1993}}}&\multicolumn{1}{c}{Labels\footnote{clump IDs from \citet{Chris2005}}}&\multicolumn{1}{c}{ }&\multicolumn{1}{c}{Diameter}&\multicolumn{1}{c}{Velocity}&\multicolumn{1}{c}{dV}&\multicolumn{1}{c}{Factor}&\multicolumn{2}{c}{Temperature Tb} \\
\multicolumn{7}{c}{ }&\multicolumn{1}{c}{Observed}&\multicolumn{1}{c}{Resolved} \\
\multicolumn{1}{c}{ }&\multicolumn{1}{c}{ }&\multicolumn{1}{c}{ }&\multicolumn{1}{c}{pc}&\multicolumn{1}{c}{kms$^{-1}$}&\multicolumn{1}{c}{kms$^{-1}$}&\multicolumn{1}{c}{ }&\multicolumn{1}{c}{K}&\multicolumn{1}{c}{K} \\
\hline
{A}&{D}&{H$^{12}$CN}&0.17&102&50&0.093&3.6&39 \\
 & &{HCO$^{+}$}& & & & &3.4&37 \\
 & &{H$^{13}$CN}& & & & &0.8&9 \\
 & &{HCN 3-2}& & & & &4.4&47 \\
 \\
 {B}&{W}&{H$^{12}$CN}&0.27&65&60&0.234&4.8&21 \\
 & &{HCO$^{+}$}& & & & &3.4&14.5 \\
 & &{H$^{13}$CN}& & & & &1.3&5.6 \\
% & &{HCN 3-2}& & & & & & \\
 \\
 {C}&{Z}&{H$^{12}$CN}&0.24&55&60&0.185&2.9&16 \\
 & &{HCO$^{+}$}& & & & &3.0&16 \\
 & &{H$^{13}$CN}& & & & &1.3&7 \\
 & &{HCN 3-2}& & & & &2.5&13.5 \\
 \\
 {D}&{na}&{H$^{12}$CN}&0.25&85&75&0.200&3.0&15 \\
  & &{HCO$^{+}$}& & & & &{$\leq$1.0}&5 \\
  & &{H$^{13}$CN}& & & & &0.7&3.5 \\
  & &{HCN 3-2}& & & & &2.1&10.5 \\
  \\
  {E}&{O}&{H$^{12}$CN}&0.33&-96&88&0.349&4.4&12.6 \\
 & &{HCO$^{+}$}& & & & &2.2&6.3 \\
 & &{H$^{13}$CN}& & & & &0.7& 2.0 \\
 & &{HCN 3-2}& & & & &5.0&14.3 \\
\hline      
\end{tabular}

\end{minipage}
\end{table*}

\paragraph*{}
We selected the second group of seven clumps, labelled D, I, M, O, P, W and Z by \citet{Chris2005} by comparing locations and their central velocities listed in three separate papers that described observations in three HCN transitions and the HCO$^{+}$(1-0) transition \citep{Chris2005,MMC2009,Jacks1993}.  

\paragraph*{}
The velocity integrated HCN and HCO$^{+}$ fluxes from \citet{Chris2005} have an
estimated uncertainty of 10\%, HCN 4-3 integrated intensities from
\citep{MMC2009} have an estimated uncertainty of 0.4K based on the 0.5Jy cut off
for the integrated map, equating to about 10\% for the peak brightness temperatures. The HCN 3-2 data is the same as used for the first group of clumps. For each transition we estimate a further 5\% uncertainty is due to the extraction of peak brightness temperatures from the published spectra.
\paragraph*{}
Four of the seven objects, i.e.\ clumps D, M, W and Z, have the strongest velocity space correlation and are undoubtedly clumps observed in multiple HCN transitions. Clumps I and O have anomalous central velocities in the 3-2 transition. These observations have been included on the basis that the spectra are unresolved, which can leave greater room for discrepancies and were based on visual inspection of the \citet{Jacks1993} figures. Clump P has a lower central velocity in the HCN(4-3) transition, but has been included on the basis that it was one of the clumps matched by \citet{MMC2009} with the HCN(1-0) observations by \citet{Chris2005}. Fig.\ 7 in \citet{MMC2009} shows the (1-0) and (4-3) spectra with double peaks and absorption occurs between the peaks in the (1-0) spectrum which can explain the discrepancies (see Table \ref{Vels}). The larger peak intensity value was used where twin peaks occurred together with the published FWHM spectral widths. All clumps were selected to provide a representative sample through the ring.

\paragraph*{•}
It should be noted that central velocities were only quantified by \citet{Chris2005} in Table 2 of their paper for the (1-0) transition. Central velocities for the (3-2) and (4-3) transitions were estimated from the spectra in the relevant papers. Line widths were listed for the (1-0) and (4-3) transitions but had to be estimated for the (3-2) transition. The (3-2) transition spectral widths are large, due in some measure to the larger telescope beam size (12$^{''}$).
\begin{table*}
\begin{minipage}{145mm}
\small
\caption{\label{Vels}  Properties of 2nd Clump Group} 
\begin{tabular}{cccccccccc}
\hline
\multicolumn{2}{c}{HCN clump}&\multicolumn{3}{c}{Central Velocity}&\multicolumn{3}{c}{FWHM}&\multicolumn{2}{c}{Position rel to SgrA$^{*}$} \\
\multicolumn{2}{c}{ }&\multicolumn{3}{c}{km s$^{-1}$}&\multicolumn{3}{c}{km s$^{-1}$}&\multicolumn{1}{c}{$\Delta$X}&\multicolumn{1}{c}{$\Delta$Y} \\
\multicolumn{1}{c}{(1-0)\footnote{clump IDs from \citet{Chris2005}}}&\multicolumn{1}{c}{(4-3)\footnote{clump IDs from \citet{MMC2009}}} &\multicolumn{1}{c}{(1-0)}&\multicolumn{1}{c}{(3-2)}&\multicolumn{1}{c}{(4-3)\footnote{Estimates from spectra presented in papers}}&\multicolumn{1}{c}{(1-0)}&\multicolumn{1}{c}{(3-2)$^{d}$}&\multicolumn{1}{c}{(4-3)} & \multicolumn{1}{c}{pc}&\multicolumn{1}{c}{pc}  \\
\hline
{D}&{A}&101&100&110&45.5&80.0&38.5&1.09&1.32 \\
{I}&{AA}&-18&-50&-25&15.0&80.0&49.5&0.83&-0.38 \\
{M}&{U}&-64&-50&-50&19.0&75.0&47.0&-0.21&-1.49 \\
{O}&{Q}&-108&-70&-90&36.5&90.0&51.0&-0.81&-1.41 \\
{P}&{N}&-73&-75&-40&28.2&75.0&97.0&-0.81&-0.86 \\
{W}&{F}&56&45&50&27.9&45.0&40.0&-0.33&0.88 \\ 
{Z}&{E}&58&50&40&39.6&50.0&55.0&-0.10&1.54 \\
\hline
\end{tabular}

\end{minipage}
\end{table*}

\begin{figure} 
\includegraphics[scale=0.50,bb=  75 170 510 670]
{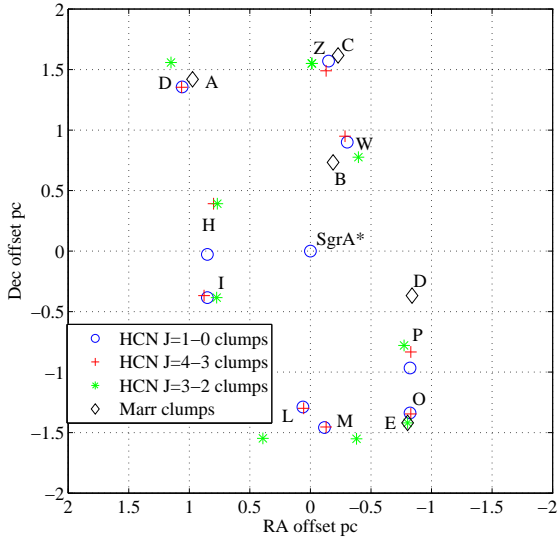} \fontsize{9} {9} 
\caption[Plot of HCN Clumps Selected for LVG Modelling]{\label{Coclumps}  The figure shows positions of the two HCN clump groups selected for LVG modelling. Locations of the first group are marked with black diamonds and have black labels used for \citet{Marr1993} clumps. The location of the second clump group transitions are marked with their respective symbols as shown in the legend and have blue labels used for \citet{Chris2005} clumps. Offsets from SgrA$^*$ are in parsecs based on the J2000 epoch.} 
\end{figure}

\paragraph*{} 
Table \ref{Vels} and Fig.\ \ref{Coclumps} show nine clumps with the observations in three HCN transitions located in reasonable proximity of one another together with the offsets from SgrA$^{*}$ in parsecs for the transition observations. Seven of these nine clumps (D, I, M, O, P, W and Z) have their different transition observations within 7 arcseconds and generally within 2.6 arcsecs of their mean location and the emission can be regarded as arising from the same clump especially given that the uncertainty of the positions of the HCN(3-2) clumps are $\pm 0.25$pc or 6.5 arcsecs due to the telescope's beam size. Clumps H and L have the (1-0) and (4-3) transitions in close proximity but the (3-2) transition is too distant to be from the same clump. For convenience the clumps chosen for analysis are referred to by the labels given in \citet{Chris2005}. 

\paragraph*{}
Four clumps (A, B, C and E) from the first group are in common with clumps (D, W, Z and O) of the second  group, allowing a comparison between the results. Our results and conclusions are based on observations of the second group of clumps which are largely resolved.

\section{LVG Modelling} \label{LVG}
In this section, we use a Large Velocity Gradient model to simultaneously fit HCN and HCO$^{+}$ peak line intensities in order to infer hydrogen number densities, optical depths and the column densities of HCN and HCO$^{+}$ in the CND clumps. Data for the two HCN clump groups selected from Section \ref{clumpselect} are analysed. We adopted an ortho to para hydrogen  ratio of 3:1 as well as a [He]/[H] ratio of 0.1. 

\paragraph*{}
Molecular data for HCN \citep{Green1974} and HCO$^{+}$ \citep{Flower1999} was sourced from the Leiden Atomic and Molecular Database (LAMDA)\footnote {URL http//:www.strw.leidenuniv.nl/\~{} moldata}. This data included excitation level information, Einstein A coefficients and collision rates for a range of kinetic temperatures and formed part of the input to the LVG code  used to analyse the observations. We do not use the HCN database with hyperfine lines in the 1-0 transition as the high degree of overlap of the hyperfine spectra combined with the large velocity width of the observations produce no significant difference from the data set using the single 1-0 transition.    

\paragraph*{}
Electron collisions with HCN were not included in our analysis because the estimated low electron fraction of $\leqslant$ 10$^{-7}$, corresponding to a hydrogen molecular density of 10$^{6}$cm$^{-3}$, would contribute only a small percentage of the total intensity. Given the high column density of the cloud, the ionisation fraction is not expected to exceed 10$^{-7}$ whereas significant effects would require the fraction to be $\sim$10$^{-5}$.
Figures 8a \& b from \citet{Sternberg1995} support this assumption of low ionisation fraction for galactic centre clouds which have an Av of $\sim$27. The cosmic ray ionisation rate would need to be in excess of $5\times10^{-13}$\,s$^{-1}$\,H$^{-1}$ in order to raise the electron abundance to the point that electron collisions are competitive. This is unlikely but may be achievable in the close vicinity of a supernova shock wave, and the circumnuclear ring may be disturbed by the expansion of the Sgr A East supernova remnant \citep{FYZ2001,Lee2008}. 
\paragraph*{}
We checked our LVG code with the RADEX code which uses the same molecular database for less than a 0.5\% difference in results, which shows a strong agreement between the two models, given there are presumably differences in radiation fields and allowances for dust emission.  

\paragraph*{}
150\,K was chosen as the fiducial value for the gas temperature for modelling as this was midway between 250\,K, chosen by \citet{Marr1993}  and 50\,K, used by \citet{Chris2005}. We look at the effect of varying the kinetic temperature in Sections \ref{Marr} and \ref{Chris}.

\paragraph*{}
 \citet{Chris2005} quoted dust temperatures varying from 20 to 80\,K and adopted 50\,K and an Av extinction of 30 in their calculations. \citet{Genzel1989} quoted dust temperatures between 50 and 90\,K. \citet{Mezger1989} indicates that cold dust (20\,K) within the CND accounts for $\sim$ 90\% of the 1.3mm dust emission \citet{Wade1987} found a uniform  extinction of $\sim$ 27 towards the central 0$^{'}$.5 of the galactic centre. We adopt a dust temperature of 75K and A$_{V}$ $=$ 26.9 , which is calculated by extrapolating an observed intensity of 2$\times$10$^{11}$\,Jy at 100\,$\mu$m \citet{Becklin1982} to the equivalent intensity at 1.3\,mm  and calculating its associated optical depth which is converted using data provided by \citet{Draine1984}. We found the model proved insensitive to variation of these parameters because the emission frequencies of the molecular tracers are on the low frequency tail of the dust's IR emission spectrum.

\paragraph*{}
Intensities for the first group of clumps are expressed as resolved peak brightness temperatures obtained by applying dilution factors, which are based on on the FWHM clump diameters listed in Table 2 of \citet{Chris2005}, to the observed radiation temperatures, T$_{R}$ = I$_{\nu}$c$^{2}$/2k $\nu^{2}$, listed in Table 1 of \cite{Marr1993}. See Table \ref{Gp1} for details. 

\paragraph*{}
Peak brightness temperatures  \footnote{This paper uses brightness to mean radiation temperature as defined above not based on the Planck function} for the second group were derived by applying the Rayleigh Jeans formula to integrated intensities from \citet{Chris2005} for the 1-0  and \citet{MMC2009} for the 4-3 transition. Peak brightness temperatures for the 3-2 transition were estimated from spectra in \citet{Jacks1993} and corrected with a dilution factor of 0.2, based on an average FWHM clump size of $6^{''}.45$ and a telescope beam size of 12$^{''}$ $\times$ 12$^{''}$.
 
\paragraph*{}
Each molecule was modelled by starting with local thermal equilibrium (LTE) conditions by choosing a sufficiently high hydrogen number density (10$^{12}$ cm$^{-3}$), then for each increasing value (dex 0.1) of HCN column density from 10$^{12}$ to 10$^{18}$ cm$^{-2}$ the hydrogen number density was decreased in steps of (dex 0.1) to 10$^{3}$ cm$^{-3}$. The model intensities were plotted as contours on a log-log plot of HCN column density per unit line width at FWHM as abscissa, Ncol/dV, ( cm$^{-2}$/(km s$^{-1}$)) and hydrogen number density as ordinate, n(H) (cm$^{-3}$) . The hydrogen number density and the HCN 1-0 column density per unit line width were inferred from the average of the intersection points of the brightness contours of the four molecular tracers plotted for each clump. The tracers for Group 1 were H$^{12}$CN, H$^{13}$CN , HCO$^{+}$ and HCN 3-2 while HCN 1-0, HCN 3-2, HCN 4-3 and HCO$^{+}$ 1-0 were used for Group 2.

\section{Results}\label{Results}

\subsection{Group 1 Clumps} \label{Marr}
  
\paragraph*{}
Table 3 summarises the results from modelling the Group 1 clumps based on observations of H$^{12}$CN 1-0  by \citet{Guesten1987}, H$^{13}$CN 1-0 and HCO$^{+}$ 1-0 by\citet{Marr1993} and HCN 3-2  by \citet{Jacks1993}.  
The hydrogen densities varied from 0.5 to 0.9$\times10^{6}$ cm$^{-3}$ and HCN
column per unit line width varied from 0.3 to 1.3$\times10^{14}$
cm$^{-2}$kms$^{-1}$. Corresponding densities were also recorded for the
intersection of the H$^{12}$CN and H$^{13}$CN intensity contours which produced
lower hydrogen densities (0.2 to 0.4$\times10^{6}$ cm$^{-3}$) and higher HCN
column per line densities (0.3 to 1.8$\times10^{14}$ cm$^{-2}$kms$^{-1}$). Table
4 lists the derived parameters for clump A, at the intersection of the H$^{12}$CN and H$^{13}$CN intensity contours for both 150 and 250K gas temperature. The  sensitivity to gas temperature shows excitation temperatures at 250\,K increased for H$^{12}$CN (190\%),HCO$^{+}$ (30\%)and  HCN 3-2 (105\%) but decreased for H$^{13}$CN (24\%), while optical depths decreased for all tracers H$^{12}$CN (80\%), H$^{13}$CN (17\%), HCO$^{+}$ (31\%) and  HCN 3-2 (180\%).

\paragraph*{}

\begin{figure} 
\includegraphics[scale=0.52,bb=  0 40 455 700] 
{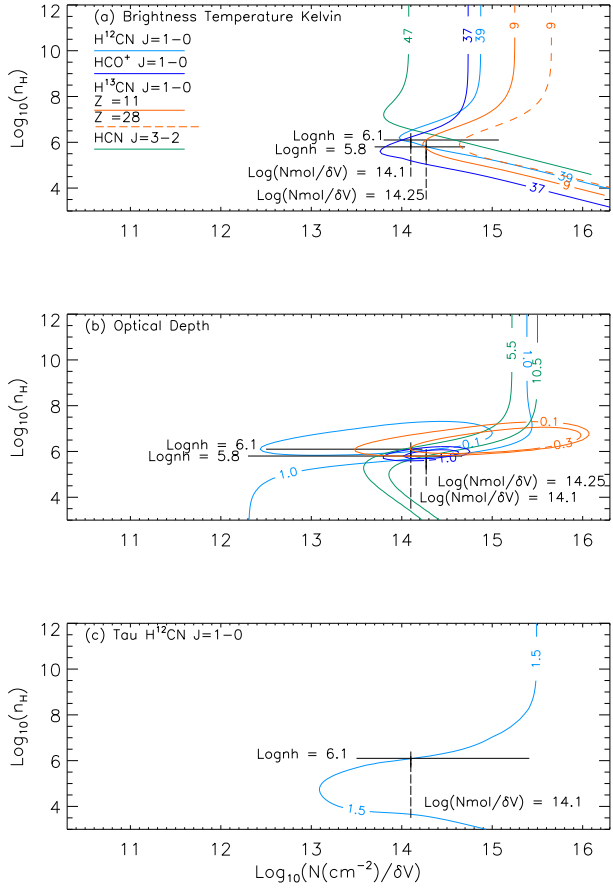} \fontsize{9} {9} \vspace{-0.5cm} 
\caption[Marr clump A Tb and Optical Depth Plots]{\label{MarrA} LVG results for Clump A of the first group. The lesser abundant HCO$^{+}$ and H$^{13}$CN, were modelled over a lower range of column densities and plotting consistency with  H$^{12}$CN plots maintained by multiplying their column densities per line width  by the appropriate abundance ratio to convert them to equivalent H$^{12}$CN  column densities which were then divided by their relevant line width. \\
Panel (a) shows brightness temperature contours for H$^{12}$CN, H$^{13}$CN, HCO$^{+}$ all (1-0) and HCN(3-2). H$^{13}$CN with Z$_{C}$ = 7 (chain dotted) and Z$_{C}$ = 28 (dashed) and HCO$^{+}$ with Z$_{O^{+}}$ = 0.4  (dashed) contours are also plotted. The average intersection point of the species is logn$_{\mathrm{h}}$ = 6.25, log(N$_{\mathrm{mol}}$/dV) = 14.15 with the intersection of the H$^{12}$/H$^{13}$ contours at log(n$_{\mathrm{h}}$) = 5.9, logN$_{\mathrm{Mol}}$/dV) = 14.25.
Panel (b) shows optical depth contours for the tracers. One abundance for both H$^{13}$CN and HCO$^{+}$ are plotted as indicated for clarity. H$^{12}$CN, H$^{13}$CN and HCO$^{+}$ are optically thin with inverted transition populations, and HCN(3-2) is optically thick with \large{$\tau$} \normalsize $\sim$ 1.3.
Panel (c) shows the optical depth contours of H$^{12}$CN calculated using brightness temperatures for H$^{12}$CN and H$^{13}$CN from Molex and a [$^{12}$C]/[$^{13}$C] abundance ratio, Z = 11. \large{$\tau^{12}$} \normalsize $\sim$ 1 at the average intersection point and $\sim$ 2.0 for the closest point of approach of the H$^{12}$CN \& H$^{13}$CN brightness contours.} 
\end{figure}

The effect of varying the [$^{12}$C/$^{13}$C] ratio, Z$_{C}$, is shown in Fig.\ \ref{MarrA} where the brightness contour for H$^{13}$CN shifts to the left with decreasing Z$_{C}$ values. At Z$_{C}$ = 7 the H$^{13}$CN brightness temperature contour passes through the average  of the intersection points and indicates that this is a more representative Z$_{C}$ value  for conditions in the CND than the range of 11 to 28 adapted by \citet{Marr1993}.

\paragraph*{}
Table 4 lists the model parameters for clump A, at the intersection of the H$^{12}$CN and H$^{13}$CN brightness contours. Fig.\ \ref{MarrA} Panel(a): for a gas temperature of T = 150\,K shows this point occurs at a lower hydrogen density, higher HCN column density per line width and larger optical depths than the average values obtained for the average intersection point of the HCN transitions,  listed in Table 3. Similar values are evident for a gas temperature of 250\,K. At these points the molecular hydrogen number density n(H2) is typically 0.4 to 0.25$\times$ 10$^{6}$cm$^{-3}$, which about a half to a fifth of the densities for the average of intersection point values obtained above and the density of 10$^{6}$cm$^{-3}$ inferred by \citet{Marr1993}.

\paragraph*{}
The optical depths at the average point of intersection of individual trace
molecules produced optically thin conditions for the 1-0 transitions of
H$^{12}$CN, H$^{13}$CN and HCO$^{+}$ and optically thick conditions for the 3-2 transition of HCN. Further the 1-0 transitions of H$^{12}$CN (-0.3 to -0.9) and HCO$^{+}$ (-0.3 to -1.1) were weakly inverted, the optical depth of the 1-0 transition of H$^{13}$CN (-0.3) for Clump A is also weakly inverted.

\paragraph*{}
Panel (c) of Fig.\ \ref{MarrA} for Clump A  shows contours of H$^{12}$CN opacity calculated using H$^{12}$CN and H$^{13}$CN  peak brightness temperatures from Molex substituted into Eqn \ref{H12CN} and produces optical depths ranging from 1.0 to 2.0 for a [C$^{12}$]/[C$^{13}$] ratio Z$_{C}$ = 11, which is the minimum ratio value considered by \citet{Marr1993}. They proposed that a consistent model for CND clumps of $\approx 0.1$pc diameter and [HCN]/[H$_{2}$] ratio $\approx 0.8\times 10^{-8}$ required assuming values for HCN 1-0 optical depth of 4, a gas temperature of 250\,K and a molecular hydrogen density of $\approx$2.6 $\times$ 10$^{6}$ cm$^{-3}$. Eqn \ref{H12CN} For clump A, produces a H$^{12}$CN optical depth of 3.4 (150K)and 3.0 (250K)  (see Table 4) compared with their value for optical thickness of 4.

\clearpage
\begin{table*}
\vbox to110mm{\vfil \caption{Insert Table 3 generated from landscape tables 3 4.tex} \vfil}
\label{Marrcprops}
\end{table*} 

\begin{table*}
\vbox to110mm{\vfil \caption{Insert Table 4 generated from landscape tables 3 4.tex} \vfil}
\label{Marrcparam}
\end{table*} 

\newpage

\subsection{Group 2 Clumps} \label{Chris}

\paragraph*{}
The results for these seven clumps, labelled D, I, M, N, O, P, W and Z after \citet{Chris2005}, were obtained by averaging the values of intersection points of pairs of species and are summarised in Table 5. The molecular hydrogen densities range from n(H$\small{2}$) \normalsize  0.8 (clump I) to 1.6 (clump M) $\times10^{6}$cm$^{-3}$ for a gas temperature of 150\,K; these are less than a quarter to a half of the values for the optically thin scenario of \citet{Chris2005} for the HCN(1-0) transition. Modelling clumps D and M with a gas temperature of 50\,K produces a density, n(H$\small{2}$), \normalsize of 4.0$\times10^{6}$cm$^{-3}$ (see Table 5). 

\paragraph*{} 
Table 5 summarises the results for optical depth for the seven clumps where the HCN(1-0) and HCO$^{+}$(1-0) transitions indicate the presence of stimulated emission with negative excitation temperatures and opacities ranging from $-$0.5 to $-$0.25 for 150\,K and about 0 for 50\,K. The HCN(3-2) opacities  range from 1.1 to 2.0 and for HCN(4-3) from 1.0 to 2.6. 

\paragraph*{}
The HCN column and molecular hydrogen number densities were obtained from averaging the values of the intersection points of the HCN(1-0) brightness contour with each of the other contours. For example in Fig.\ \ref{clumpW} for clump W the intersection of HCN(1-0), HCN(4-3) and HCO$^{+}$(1-0) was the lower point at logN${\mathrm{col}}$/$\delta$V = 13.2 and logn$_{\mathrm{h}}$ = 6.1, with the intersection of HCN(1-0) and (3-2) the upper point at log(N$_{\mathrm{mol}}$/dV) = 13.4 and logn$_{\mathrm{H}}$ = 6.6. The result is an average Column density/$\delta$V of 2 $\times$ 10$^{13}$ cm$^{-2}$ and hydrogen density of 0.9 $\times$ 10$^{6}$ cm$^{-3}$.  It should be noted that the lower limit curve of 12.6\,K for the (3-2) transition provides a tighter intersection envelope and slightly lower values for both  HCN column and hydrogen number densities.  

\section{Analysis} \label{anal}
\paragraph*{}
In the first group of five clumps, \citet{Marr1993} used a statistical equilibrium excitation model for HCN with values for gas temperature ranging from 150 to 450\,K, a dust temperature of 75\,K, optical depths from 1 to 12 and [HCN]/[H$_{2}$] ratios between 6$\times$ 10$^{-9}$ and 3$\times$10$^{-6}$. The [$^{12}$C]/[$^{13}$C] ratio (Z$_{C}$) was assumed to vary between 10 and 40 while noting that previous estimates varied from 11 (in SgrB2 \citep{Magnum1988}) to 28 ( in the galactic centre \citep{Wannier1989}). 

\paragraph*{}
Fig.\ \ref{MarrA} Panel (a) shows  brightness contours for clump A with alternative abundance ratios Z$_{C}$ $\equiv$ [H$^{12}$CN]/[H$^{13}$CN] = 7 and 28 and Z$_{O}$ $\equiv$ [HCO$^{+}$]/[HCN] = 0.4 as dashed contours.
The alternative ratios of Z$_{C}$ = 28 and Z$_{O^{+}}$ = 0.4 do not provide a consistent intersection with the H$^{12}$CN contour. The contours for Z$_{C}$ = 7 and Z$_{O^{+}}$ = 0.4 do intersect the H$^{12}$CN contour but at a lower hydrogen density value than for Z$_{C}$ = 11 and Z$_{O^{+}}$ = 0.74 which led to the adoption of Z = 11$_{C}$ and Z$_{O^{+}}$ = 0.74 for the other clump plots. It should also be noted that the intersection area for all tracers is not representative of the intersection points of the H$^{12}$CN and H$^{13}$CN brightness contours, which occur at higher column density per line width (NCol$\delta$V) values ranging from 0.3 to 1.8$\times 10^{14}$ cm$^{-2}$kms$^{-1}$ and lower Hydrogen density values of 0.13 to 0.4$\times$10$^{6}$cm$^{-3}$.

\paragraph*{}
The effect of changing the Z$_{C}$ abundance ratio is shown in Fig.\ \ref{MarrA} Panel(a) where the relevant brightness contours shift to the left as the ratio decreases. This is especially noticeable for the [C$^{12}$]/[C$^{13}$] = 7 curve where the contour falls within the intersection points of the other three species. Similarly, a reduction in the brightness temperature shifts the contour to the left.

\paragraph*{}
The  second group of seven HCN clumps provided a good representative sample of CND conditions as they are spread throughout the ring. Clump D is in the North East, clump I is in the East, clump M is in the South, clumps O \& P are in the South West and clumps W \& Z are in the North (see Fig.\ \ref{Coclumps}).

\paragraph*{}
The accuracy of the HCN(3-2) peak brightness temperatures is stated by \citet{Jacks1993} to be $\pm 30\%$ due to the relatively large beam size of about 12'', compared to the average clump size of 6.5'', this also contributed to the large FWHM line widths for this transition compared to the other transitions of HCN and HCO$^{+}$ for this group of clumps (see column 7 of Table \ref{Vels}). The effect of the uncertainty can be seen in the plot of peak brightness temperatures for clump W in Fig.\ \ref{clumpW}, where the contour for the lower peak brightness temperature of 14.7K more closely approaches the average intersection point for the other transitions than does the average HCN(3-2) peak brightness temperature contour of 21K.   

\begin{figure} 
%\vspace{110mm}
\includegraphics[scale=0.52,bb=  0 20 455 700] 
{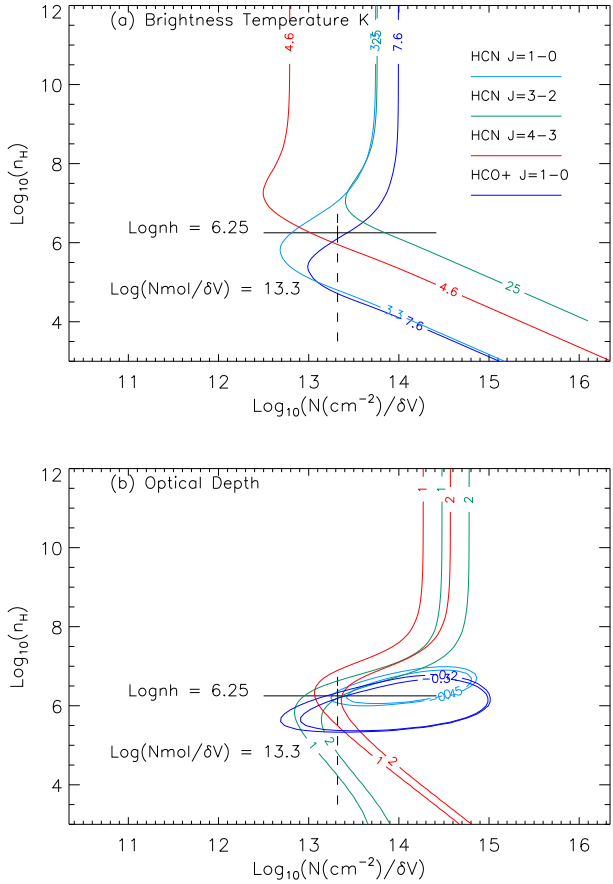} \fontsize{9} {9} \vspace{-1cm} 
\caption[clump D Tb and Optical Depth Plots]{\label{clumpD} A sample of output for clump D of the second group with gas temperature of 150\,K.  Panel (a) shows   peak brightness temperature contours for HCN (1-0), (3-2), (4-3) and HCO$^{+}$ (1-0). The arithmetic average of the species intersection points is log(n$_{H}$) = 6.25, log(NMol/dV) = 13.3.
Panel (b) shows the optical depth contours for the HCN(1-0) and HCO$^{+}$(1-0) tracers are optically thin with  inverted populations, while the HCN (3-2) and (4-3) transitions are optically thick.}
\end{figure}

\begin{figure} 
%\vspace{110mm}
\includegraphics[scale=0.52,bb=  0 20 455 700] 
{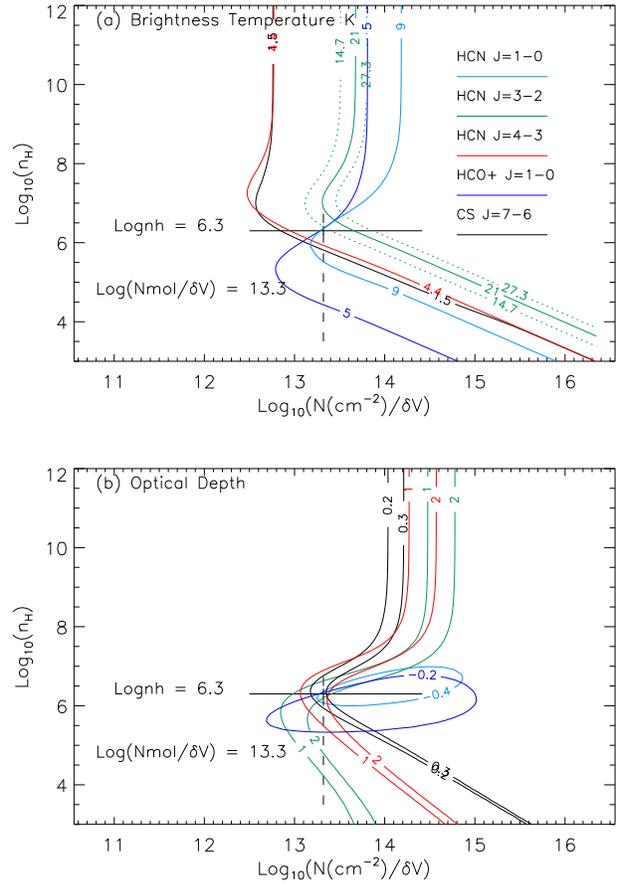} \fontsize{9} {9} \vspace{-1cm} 
\caption[clump W Tb and Optical Depth Plots]{\label{clumpW} As for Fig.\ \ref{clumpD} but for clump W. The $\pm 30\%$ uncertainty in (3-2) peak brightness temperatures is shown with the dashed contours in panel(a). CS(7-6) has also been plotted and is consistent for the same physical conditions as HCN(4-3). Panel(b) shows the CS(7-6) is optically thin in contrast with the HCN(4-3) which is optically thick.} 
\end{figure}

\paragraph*{}
Fig.\ \ref{clumpD} for clump D panel (a) shows contours for three values of the $ [\mathrm{HCO^{+}}]/[\mathrm{HCN}] $ abundance ratios, Z$_{O^{+}}$ = 0.4, 0.74  and 1.2, these variations in the [HCO$^{+}$]/[HCN] emission and absorption ratios are attributable to greater abundances of HCO$^{+}$ in lower density regions both within the CND and along the line of sight \citep{Chris2005}. Lowering the abundance ratio reduces the column density per line width value. Again in modelling the second group of HCN clumps, the mean HCO$^{+}$ abundance value of 0.74 was used in all plots, in contrast to the value of 0.4 used by \citet{Chris2005} for regions in the CND with relatively high HCN emission and low HCO$^{+}$ emission.

\paragraph*{}
The molecular hydrogen number density for the seven HCN clumps varies from 1 to 2 $\times10^{6}$cm$^{-3}$, which is an order of magnitude lower than the value n(H$_{2}$) of 10$^{7}$cm$^{-3}$ reported by \citet{Chris2005} for HCN(1-0) optically thick gas with ${\tau}$ = 4. Clumps D and M were also modelled using a gas temperature of 50\,K, which  produced molecular hydrogen densities for both of  4$\times10^{6}$ cm$^{-3}$, which agreed closely with the \citet{Chris2005} values for optically thin conditions. The brightness temperature value of the HCN(3-2) transition contours in clumps D and O was varied to account for the different dilution factor when using the clump sizes listed in Table 2 of \citet{Chris2005} instead of the average 0.25 pc size adopted for calculating the 0.2 dilution factor which was felt appropriate for representing the size of all seven clumps. Fig.\ \ref{clumpD} shows that the HCN(3-2) brightness contour shifts to the left as the brightness temperature decreases, so that it forms a tighter intersection area with the other HCN transitions, this results in slightly lower molecular hydrogen number and HCN column per line width densities than the densities in Table 5.    

\clearpage
\begin{table*}
\vbox to220mm{\vfil \caption{Insert Table 5 generated by landscape tables 5.tex} \vfil}
\label{Marrcprops}
\end{table*} 
\newpage

\section{Discussion} \label{discuss}
\paragraph*{}
Our LVG modelling shows that molecular hydrogen densities are about 10$^{6}$cm$^{-3}$ and optical depths for the HCN(1-0) transition are $\ll$ 1. These results are in marked contrast to the conclusions of \citet{Marr1993} who argue convincingly that an optical depth of 4 in HCN 1-0 leads to higher hydrogen number densities.

\paragraph*{}
Given the errors in flux obtained from the source papers are some 10 to
30\% which combined with the systematic errors arising from the calibration of
the use of different telescopes ensures that the derived densities could vary by
up to a factor of two, so that n(H$_{2}$) could range from 0.5 to 1.5 $\times$
10$^{6}$cm$^{-3}$.

\paragraph*{}
As a check on the derived physical conditions the CS(7-6) \citet{MMC2009} was modelled for Clump F,which corresponds to Clump W for HCN(1-0). The CS( 7-6) emission defines the same physical conditions as the HCN(4-3) for this clump while the CS(7-6) opacity is thin (0.2) in contrast to an optically thick (2) for HCN(4-3) (see Fig \ref{clumpW}).  

\paragraph*{}
\citet{Marr1993} argued that if both H$^{12}$CN and H$^{13}$CN occupy the same space, their emissions would have similar beam filling factors and background emissions. Then given equal excitation temperatures for the two species and optical depths that preclude enhanced H$^{12}$CN relative to H$^{13}$CN emission then the ratios of both intensities and opacities would be equal to the $^{12}$C/$^{13}$C abundance ratio, Z$_{C}$. 

%\\
%\paragraph*{}
%For $\tau _{13}$/{$\tau _{12}$ $\ll$ 1
\begin{equation}\label{T13T12}
\frac{T_{13}}{T_{12}} = \frac{(1-\mathrm{exp}(-\tau _{13}))}{(1-\mathrm{exp}(-\tau _{12}))} = \frac{\tau _{13}}{\tau _{12}} \quad for \quad \tau _{13}/\tau _{12} \ll 1     \\
\end{equation}
where the brightness temperature, T =$\lambda^{\!2}$ I$_{\nu}$/{2k}, I$_{\nu}$ is the line's peak intensity and the subscripts 12 and 13 refer to H$^{12}$CN and H$^{13}$CN respectively.
In the limit of small opacities for both species Eqn \ref{T13T12} becomes,
\begin{equation} \label{Tb1213}
\frac{T_{12}}{T_{13}} =  \frac{[\mathrm{H^{12}CN}]}{[\mathrm{H^{13}CN}]} = Z_{C} \,.
\end{equation}
   
\paragraph*{}
They then argue that the observed H$^{13}$CN emission is optically thin and so the H$^{12}$CN emission is optically thick with the H$^{12}$CN opacity, \large{$\tau$}\normalsize(12), calculated as follows,  
\begin{equation} \label{H12CN}
 \large{\tau}\normalsize(12) = -Z\ln\left[ 1-T(13)/T(12) \right]  \,,
\end{equation}

This argument hinges on the the assumption that the excitation temperatures of H$^{12}$CN and H$^{13}$CN are similar. However, excitation temperature is very sensitive to the relative populations of the upper and lower states for weakly inverted transitions. The ratio of population levels is given by

\begin{equation}
\frac{n_{2}}{n_{1}} = \frac{g_{2}}{g_{1}}\exp \left(-\frac{h\nu}{\mathrm{kT}_{ex}}\right) \, , \\
\end{equation}
then 

\begin{eqnarray}
\frac{\mathrm{d}\ln(T_{ex})}{\mathrm{d}\ln(n_{2}/n_{1})} &=& \frac{kT_{ex}}{h\nu} =\frac{T_{ex}}{4.25\,\mathrm{K}} \, 
\end{eqnarray}
\\
for the HCN(1-0) transition. When the absolute value of T$_{ex}$ is large, as it may be for weakly or nearly inverted transitions, small changes in the populations of the upper and lower states can lead to large differences of excitation temperatures.

\begin{figure} 
%\vspace{80mm}
\includegraphics[scale=0.48,bb=  0 150 505 640] 
{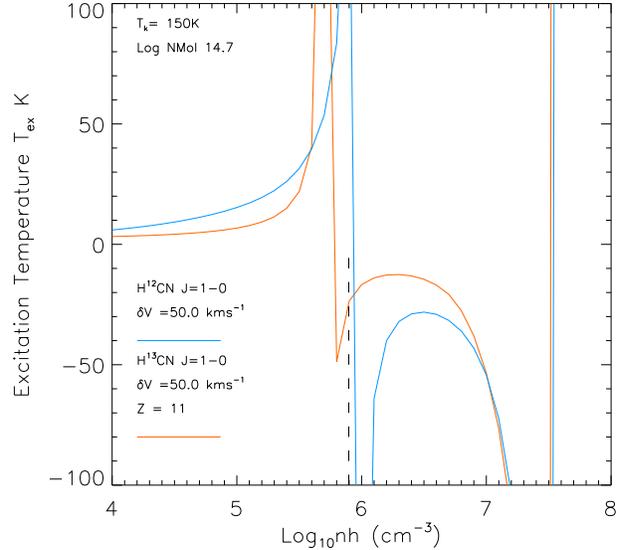} \fontsize{9} {9} 
\caption[Marr clump A Excitation Temperature vs Log Hydrogen Density]{\label{H12H13Tx}  shows the dependence of excitation temperature for the (1-0) transitions of H$^{12}$CN \& H$^{13}$CN (1-0) on Hydrogen number density at a column density of 10$^{14.7}$. The vertical dashed line at 10$^{5.9}$ corresponds to the inferred density for clump A.} 
\end{figure}

\paragraph*{}
Table 3 shows excitation temperatures for H$^{12}$CN(1-0) varying between $-$11 and $-$27.6\,K  for the (1-0) transition   and from 5.23 to 19.6\,K for H$^{13}$CN. Thus small changes in n$_{2}$/n$_{1}$ lead to large changes in T$_{\mathrm{ex}}$. By way of example Fig.\ \ref{H12H13Tx}  shows the H$^{12}$CN and H$^{13}$CN excitation temperatures for (logN$_{\mathrm{HCN}(1-0)}$ = 14.7) depend on hydrogen density. The H$^{12}$CN and  H$^{13}$CN excitation temperatures are clearly disparate when the lines are weakly or nearly inverted.

\paragraph*{}
\citet{Marr1993} also argued that because HCN and HCO$^{+}$ have similar energy level structures and trace the same gas in the CND the ratio of their abundances is equal to the ratio of their opacities times the ratio of their Einstein A coefficients.  

\begin{equation}\label{HCO+HCN}
\frac{[\mathrm{HCO}^{+}]}{[\mathrm{H^{12}CN}]} =  \frac{[\tau \mathrm{HCO}^{+})]}{[\tau \mathrm{H^{12}CN}]} \times \frac{\mathrm{A}_{21}(\mathrm{H^{12}CN})}{\mathrm{A}_{21}(\mathrm{HCO}^{+})} = 0.44 \times \frac{[\tau \mathrm{HCO}^{+}]}{[\tau \mathrm{H^{12}CN}]} \,. 
\end{equation}

Using Eqn \ref{HCO+HCN} and the optical depth values from our model for $\tau$ HCO$^{+}$ and $\tau$H$^{12}$CN in Table \ref{Marrcparam}, the [HCO$^{+}$]/[HCN] ratio ranges between 0.4-0.6. 

\paragraph*{}
\citet{Marr1993} summarised their results as [HCO$^{+}$]/[{H$^{12}$CN}] ratio varying between 0.06-0.1, with a [C$^{12}$]/[C$^{13}$] value of Z$_{C}$ = 30. These [HCO$^{+}$]/[{H$^{12}$CN}] values were significantly lower than the rates determined from higher resolution observations as between 0.4 and 1.1 \citep{Chris2005}. Our model used values [HCO$^{+}$]/[H$^{12}$CN] = 0.74 and [H$^{12}$CN]/[H$^{13}$CN] = 11  and produced more consistent intersection points for the peak brightness temperature contour plots of the respective molecules than the abundance values adopted in papers by \citet{Marr1993} and \citet{Chris2005}.

\paragraph*{}
Our modelling produced molecular hydrogen densities from 0.13 and 0.63 $\times 10^{6}$cm$^{-3}$ for the first clump group and 1.0 and 2.0 $\times 10^{6}$cm$^{-3}$ for the second group at a gas temperature of T = 150\,K. Modelling cooler gas, T = 50\,K, produced higher densities of 4 $\times 10^{6}$cm$^{-3}$, which is consistent with cooler gas being denser but not as dense as proposed by \citet{Chris2005} who used three scenarios for predicting hydrogen density viz optically thin, virial and optically thick ($\tau$ = 4) without LVG modelling. Our results agree with the optically thin scenario which implies that the clumps are tidally unstable. 

\paragraph*{}
Table \ref{[HCN]/[H2]} presents values for the [HCN]/[H$_{2}$] ratio based on hydrogen densities and HCN (1-0) column densities per unit line width obtained from our LVG modelling. The HCN densities were calculated by multiplying the HCN column densities per unit line by the appropriate FWHM line width,(dV), and then dividing the resultant HCN column density by the clump size. The [HCN]/[H$_{2}$] abundance for the first group was 4.1 $\times10^{-9}$ and for the second group abundances ranged from 0.2 and 1.2 $\times10^{-9}$ which is comparable with 10$^{-9}$ used by \citet{Chris2005}. Clumps A and E of the first group are the same as clumps D and O of the second group with the higher abundance ratio for the first group resulting from the smaller unresolved clump size 0.1pc compared to the resolved clump sizes of 0.43 and 0.33pc for the second clump group.

\begin{table*}
\begin{minipage}{180mm}

\caption{\label{[HCN]/[H2]} [HCN]/[H$_{2}]$ Abundance Ratios from Modelled Clumps}
\scriptsize
\begin{tabular}{ccccccccc} %\hline
\hline
\multicolumn{1}{c}{Clump}&\multicolumn{1}{c}{Kinetic}&\multicolumn{1}{c}{FWHM}&\multicolumn{1}{c}{HCN}&\multicolumn{1}{c}{Line}&\multicolumn{1}{c}{HCN}&\multicolumn{1}{c}{HCN}&\multicolumn{1}{c}{Hydrogen}&\multicolumn{1}{c}{[HCN]/[H$_{2}$]} \\
\multicolumn{1}{c}{Group}&\multicolumn{1}{c}{Temp}&\multicolumn{1}{c}{Diameter}&\multicolumn{1}{c}{Column Density per}&\multicolumn{1}{c}{Width}&\multicolumn{1}{c}{Column}&\multicolumn{1}{c}{Density}&\multicolumn{1}{c}{Density}&\multicolumn{1}{c}{Abundance} \\
\multicolumn{1}{c}{ID}&\multicolumn{1}{c}{ }&\multicolumn{1}{c}{}&\multicolumn{1}{c}{line width}&\multicolumn{1}{c}{dV}&\multicolumn{1}{c}{Density}&\multicolumn{1}{c}{}&\multicolumn{1}{c}{n(H2)}&\multicolumn{1}{c}{Ratio} \\
\multicolumn{1}{c}{Pairs}&\multicolumn{1}{c}{(K)}&\multicolumn{1}{c}{(pc)}&\multicolumn{1}{c}{($\times10^{13}$cm$^{-2}$/kms$^{-1}$)}&\multicolumn{1}{c}{kms$^{-1}$}&\multicolumn{1}{c}{($\times10^{14}$cm$^{-2}$)}&\multicolumn{1}{c}{($\times10^{-4}$cm$^{-3}$)}&\multicolumn{1}{c}{($\times$10$^{6}$cm$^{-3}$)}&\multicolumn{1}{c}{($\times10^{-9}$)} \\
\hline
{First $^{1}$}& & & & & & & & \\
{A}&150&{0.10 $^{2}$}&1.0&50&5.0&16.20&0.397&4.1 \\
{E}&150&{0.10 $^{2}$}&1.0&80&8.0&2.6&0.629&4.1 \\
{Second $^{3}$}& & & & & & & & \\
{D}&150&0.43&2.0&45.5&9.08&6.83&1.26&0.5 \\
{D}&50&0.43&2.5&45.5&11.40&8.6&3.96&0.2 \\
{O}&150&0.33&3.2&36.5&8.80&11.3&1.26&0.9 \\
\hline

\multicolumn{9}{l}{[1] \citet{Marr1993} labels} \\
\multicolumn{9}{l}{[2] average clump diameter from \citet{Marr1993}} \\
\multicolumn{9}{l}{[3] \citet{Chris2005} labels} \\
\end{tabular}
\end{minipage}
\end{table*}

\paragraph*{}
More detailed results can be found in \citet{Smith2012}. 
 
\section{Conclusions} \label{concl}

\paragraph*{} 
Our paper selected two groups of clumps observed in multiple transitions of HCN and the(1-0) transition of HCO$^{+}$ that coincided spatially and kinematically. The first group had been selected by \citet{Marr1993} and our analysis is based on our remodelling of their observations. We selected the second group from three separate sources HCN 1-0 and HCO$^{+}$ 1-0 from \citet{Chris2005}, HCN 3-2 from \citet{Jacks1993} and HCN 4-3 from \citet{MMC2009}. 

\paragraph*{}
Data from twenty-six HCN (1-0) clumps listed in \citet{Chris2005} provided the information for calculating the true (deprojected) clump offsets from SgrA$^{*}$. Eighteen of the twenty-six clumps fit within the accepted range of disk parameters, while seven clumps have velocities that are significantly different from the model disk. The deprojection clearly shows the circular structure of the CND with an inner cavity of radius $\sim$ 1.6 pc. The HCN(1-0) clump line of sight velocities  demonstrates the warped nature of the disk's rotating streams and that the rotational velocity of some streams vary significantly from the accepted 110 km s$^{-1}$.  

\paragraph*{•} 
We found that n(H$_{2}$) was $\sim$ 10$^{6}$\,cm$^{-3}$ and $\tau \sim -0.1$ in HCN 1-0 consistent with dust emission and limits based on \citet{Genzel2010}. This contradicts the high densities and optical depth $\tau \sim$4 inferred by \citet{Chris2005} and \citet{MMC2009}. We show that $\tau \sim$4 is based on the assumption that T$_{ex}$ for H$^{12}$CN 1-0 and H$^{13}$CN 1-0 is the same and show that this is not the case for the conditions in the CND, where the lines are weakly inverted.

\paragraph*{•}
A wider CND survey of the H$^{13}$CN(1-0) transition  might clarify the optically thick/thin question and hence the hydrogen density value. An analysis of the NH$_{3}$ (1-1), (2-2), (3-3) and (6-6)  transitions \citep{HandH2002} would also be useful to check hydrogen densities, particularly as the (6-6) transition traces denser and hotter regions more effectively than HCN. Such a study should be done in the future to reinforce the results of this study. 

\paragraph*{•}
A possible test of  maser effects of negative optical depths in the (1-0) transitions of HCN and HCO$^{+}$ would be to search for a bright (T$_{b} \sim$ 1000\,K) source of continuum emission (such as a quasar or radio galaxy) aligned with one of the disk's clumps. A comparison of off HCN line frequency, on line frequency of clump observations would then detect an increase of 30 to 60\% in the background intensity caused from amplification by HCN(1-0) emission or up to 30\% by HCO$^{+}$(1-0) emission as it passes through the clump. 

\paragraph*{•}
Our results clearly favour the warm, low density scenario with a CND mass of  3 to 4 $\times$ 10$^{5}$ M$_{\odot}$ and tidal forces pulling the clumps apart.

\subsection*{Acknowledgements}

We thank Maria Montero-Casta{\~n}o for generously providing the co-ordinates for the HCN(4-3) clumps and Farhad Yusef-Zadeh for providing the co-ordinates of the methanol and water masers detected in the CND.\\ 

\bibliographystyle{mn2e}
\bibliography{references}

\begin{thebibliography}{}

\bibitem[\protect\citeauthoryear{{Becklin}, {Gatley} \& {Werner}}{{Becklin}
  et~al.}{1982}]{Becklin1982}
{Becklin} E.~E.,  {Gatley} I.,    {Werner} M.~W.,  1982, \apj, 258, 135

\bibitem[\protect\citeauthoryear{{Christopher}, {Scoville}, {Stolovy} \&
  {Yun}}{{Christopher} et~al.}{2005}]{Chris2005}
{Christopher} M.~H.,  {Scoville} N.~Z.,  {Stolovy} S.~R.,    {Yun} M.~S.,
  2005, \apj, 622, 346

\bibitem[\protect\citeauthoryear{{Draine} \& {Lee}}{{Draine} \&
  {Lee}}{1984}]{Draine1984}
{Draine} B.~T.,  {Lee} H.~M.,  1984, \apj, 285, 89

\bibitem[\protect\citeauthoryear{{Etxaluze}, {Smith}, {Tolls}, {Stark} \&
  {Gonzalez-Alfonso}}{{Etxaluze} et~al.}{2011}]{Etxa2011}
{Etxaluze} M.,  {Smith} H.~A.,  {Tolls} V.,  {Stark} A.~A.,
  {Gonzalez-Alfonso} E.,  2011, ArXiv e-prints 1108.0313

\bibitem[\protect\citeauthoryear{{Flower}}{{Flower}}{1999}]{Flower1999}
{Flower} D.~R.,  1999, \mnras, 305, 651

\bibitem[\protect\citeauthoryear{{Genzel}}{{Genzel}}{1989}]{Genzel1989}
{Genzel} R.,  1989, in {M.~Morris} ed., The Center of the Galaxy Vol.~136 of
  IAU Symposium, {The Circumnuclear Disk (review)}.
pp 393--405

\bibitem[\protect\citeauthoryear{{Genzel}, {Eisenhauer} \&
  {Gillessen}}{{Genzel} et~al.}{2010}]{Genzel2010}
{Genzel} R.,  {Eisenhauer} F.,    {Gillessen} S.,  2010, Reviews of Modern
  Physics, 82, 3121

\bibitem[\protect\citeauthoryear{{Ghez}, {Salim}, {Hornstein}, {Tanner}, {Lu},
  {Morris}, {Becklin} \& {Duch{\^e}ne}}{{Ghez} et~al.}{2005}]{Ghez2005}
{Ghez} A.~M.,  {Salim} S.,  {Hornstein} S.~D.,  {Tanner} A.,  {Lu} J.~R.,
  {Morris} M.,  {Becklin} E.~E.,    {Duch{\^e}ne} G.,  2005, \apj, 620, 744

\bibitem[\protect\citeauthoryear{{Green} \& {Thaddeus}}{{Green} \&
  {Thaddeus}}{1974}]{Green1974}
{Green} S.,  {Thaddeus} P.,  1974, \apj, 191, 653

\bibitem[\protect\citeauthoryear{{Guesten}, {Genzel}, {Wright}, {Jaffe},
  {Stutzki} \& {Harris}}{{Guesten} et~al.}{1987}]{Guesten1987}
{Guesten} R.,  {Genzel} R.,  {Wright} M.~C.~H.,  {Jaffe} D.~T.,  {Stutzki} J.,
    {Harris} A.~I.,  1987, \apj, 318, 124

\bibitem[\protect\citeauthoryear{{Harris}, {Jaffe}, {Silber} \&
  {Genzel}}{{Harris} et~al.}{1985}]{Harris1985}
{Harris} A.~I.,  {Jaffe} D.~T.,  {Silber} M.,    {Genzel} R.,  1985, \apjl,
  294, L93

\bibitem[\protect\citeauthoryear{{Herrnstein} \& {Ho}}{{Herrnstein} \&
  {Ho}}{2002}]{HandH2002}
{Herrnstein} R.~M.,  {Ho} P.~T.~P.,  2002, \apjl, 579, L83

\bibitem[\protect\citeauthoryear{{Jackson}, {Geis}, {Genzel}, {Harris},
  {Madden}, {Poglitsch}, {Stacey} \& {Townes}}{{Jackson}
  et~al.}{1993}]{Jacks1993}
{Jackson} J.~M.,  {Geis} N.,  {Genzel} R.,  {Harris} A.~I.,  {Madden} S.,
  {Poglitsch} A.,  {Stacey} G.~J.,    {Townes} C.~H.,  1993, \apj, 402, 173

\bibitem[\protect\citeauthoryear{{Karlsson}, {Sjouwerman}, {Sandqvist} \&
  {Whiteoak}}{{Karlsson} et~al.}{2003}]{Karlsson2003}
{Karlsson} R.,  {Sjouwerman} L.~O.,  {Sandqvist} A.,    {Whiteoak} J.~B.,
  2003, \aap, 403, 1011

\bibitem[\protect\citeauthoryear{{Lee}, {Pak}, {Choi}, {Davis}, {Geballe},
  {Herrnstein}, {Ho}, {Minh} \& {Lee}}{{Lee} et~al.}{2008}]{Lee2008}
{Lee} S.,  {Pak} S.,  {Choi} M.,  {Davis} C.~J.,  {Geballe} T.~R.,
  {Herrnstein} R.~M.,  {Ho} P.~T.~P.,  {Minh} Y.~C.,    {Lee} S.,  2008, \apj,
  674, 247

\bibitem[\protect\citeauthoryear{{Liszt}, {Burton} \& {van der Hulst}}{{Liszt}
  et~al.}{1985}]{Liszt1985}
{Liszt} H.~S.,  {Burton} W.~B.,    {van der Hulst} J.~M.,  1985, \aap, 142, 237

\bibitem[\protect\citeauthoryear{{Mangum}, {Rood}, {Wadiak} \&
  {Wilson}}{{Mangum} et~al.}{1988}]{Magnum1988}
{Mangum} J.~G.,  {Rood} R.~T.,  {Wadiak} E.~J.,    {Wilson} T.~L.,  1988, \apj,
  334, 182

\bibitem[\protect\citeauthoryear{{Marr}, {Wright} \& {Backer}}{{Marr}
  et~al.}{1993}]{Marr1993}
{Marr} J.~M.,  {Wright} M.~C.~H.,    {Backer} D.~C.,  1993, \apj, 411, 667

\bibitem[\protect\citeauthoryear{{Marshall}, {Lasenby} \& {Harris}}{{Marshall}
  et~al.}{1995}]{Marshall1995}
{Marshall} J.,  {Lasenby} A.~N.,    {Harris} A.~I.,  1995, \mnras, 277, 594

\bibitem[\protect\citeauthoryear{{Mehringer} \& {Menten}}{{Mehringer} \&
  {Menten}}{1997}]{Mehringer1997}
{Mehringer} D.~M.,  {Menten} K.~M.,  1997, \apj, 474, 346

\bibitem[\protect\citeauthoryear{{Mezger}, {Duschl} \& {Zylka}}{{Mezger}
  et~al.}{1996}]{Mezger1996}
{Mezger} P.~G.,  {Duschl} W.~J.,    {Zylka} R.,  1996, \aapr, 7, 289

\bibitem[\protect\citeauthoryear{{Mezger}, {Zylka}, {Salter}, {Wink}, {Chini},
  {Kreysa} \& {Tuffs}}{{Mezger} et~al.}{1989}]{Mezger1989}
{Mezger} P.~G.,  {Zylka} R.,  {Salter} C.~J.,  {Wink} J.~E.,  {Chini} R.,
  {Kreysa} E.,    {Tuffs} R.,  1989, \aap, 209, 337

\bibitem[\protect\citeauthoryear{{Montero-Casta{\~n}o}, {Herrnstein} \&
  {Ho}}{{Montero-Casta{\~n}o} et~al.}{2009}]{MMC2009}
{Montero-Casta{\~n}o} M.,  {Herrnstein} R.~M.,    {Ho} P.~T.~P.,  2009, \apj,
  695, 1477

\bibitem[\protect\citeauthoryear{{Oka}, {Nagai}, {Kamegai} \& {Tanaka}}{{Oka}
  et~al.}{2011}]{Oka2011}
{Oka} T.,  {Nagai} M.,  {Kamegai} K.,    {Tanaka} K.,  2011, \apj, 732, 120,1

\bibitem[\protect\citeauthoryear{{Oka}, {Nagai}, {Kamegai}, {Tanaka} \&
  {Kuboi}}{{Oka} et~al.}{2007}]{Oka2007}
{Oka} T.,  {Nagai} M.,  {Kamegai} K.,  {Tanaka} K.,    {Kuboi} N.,  2007,
  \pasj, 59, 15

\bibitem[\protect\citeauthoryear{{Pierce-Price}, {Richer}, {Greaves},
  {Holland}, {Jenness}, {Lasenby}, {White}, {Matthews}, {Ward-Thompson},
  {Dent}, {Zylka}, {Mezger}, {Hasegawa}, {Oka}, {Omont} \&
  {Gilmore}}{{Pierce-Price} et~al.}{2000}]{P-Price2000}
{Pierce-Price} D.,  {Richer} J.~S.,  {Greaves} J.~S.,  {Holland} W.~S.,
  {Jenness} T.,  {Lasenby} A.~N.,  {White} G.~J.,  {Matthews} H.~E.,
  {Ward-Thompson} D.,  {Dent} W.~R.~F.,  {Zylka} R.,  {Mezger} P.,  {Hasegawa}
  T.,  {Oka} T.,  {Omont} A.,    {Gilmore} G.,  2000, \apjl, 545, L121

\bibitem[\protect\citeauthoryear{{Serabyn}, {Guesten}, {Walmsley}, {Wink} \&
  {Zylka}}{{Serabyn} et~al.}{1986}]{SGW1986}
{Serabyn} E.,  {Guesten} R.,  {Walmsley} J.~E.,  {Wink} J.~E.,    {Zylka} R.,
  1986, \aap, 169, 85

\bibitem[\protect\citeauthoryear{{Serabyn} \& {Lacy}}{{Serabyn} \&
  {Lacy}}{1985}]{Serabyn1985}
{Serabyn} E.,  {Lacy} J.~H.,  1985, \apj, 293, 445

\bibitem[\protect\citeauthoryear{{Sjouwerman} \& {Pihlstr{\"o}m}}{{Sjouwerman}
  \& {Pihlstr{\"o}m}}{2008}]{Sjman2008}
{Sjouwerman} L.~O.,  {Pihlstr{\"o}m} Y.~M.,  2008, \apj, 681, 1287

\bibitem[\protect\citeauthoryear{{Smith}}{{Smith}}{2012}]{Smith2012}
{Smith} I.~L.,  2012, MPhil Thesis Macquarie University (ArXiv 1307.1255)

\bibitem[\protect\citeauthoryear{{Sternberg} \& {Dalgarno}}{{Sternberg} \&
  {Dalgarno}}{1995}]{Sternberg1995}
{Sternberg} A.,  {Dalgarno} A.,  1995, \apjs, 99, 565

\bibitem[\protect\citeauthoryear{{Sutton}, {Danchi}, {Jaminet} \&
  {Masson}}{{Sutton} et~al.}{1990}]{Sutton1990}
{Sutton} E.~C.,  {Danchi} W.~C.,  {Jaminet} P.~A.,    {Masson} C.~R.,  1990,
  \apj, 348, 503

\bibitem[\protect\citeauthoryear{{{\v S}ubr}, {Schovancov{\'a}} \&
  {Kroupa}}{{{\v S}ubr} et~al.}{2009}]{Subr2009}
{{\v S}ubr} L.,  {Schovancov{\'a}} J.,    {Kroupa} P.,  2009, \aap, 496, 695

\bibitem[\protect\citeauthoryear{{Wade}, {Geballe}, {Krisciunas}, {Gatley} \&
  {Bird}}{{Wade} et~al.}{1987}]{Wade1987}
{Wade} R.,  {Geballe} T.~R.,  {Krisciunas} K.,  {Gatley} I.,    {Bird} M.~C.,
  1987, \apj, 320, 570

\bibitem[\protect\citeauthoryear{{Wannier}}{{Wannier}}{1989}]{Wannier1989}
{Wannier} P.~G.,  1989, in {M.~Morris} ed., The Center of the Galaxy Vol.~136
  of IAU Symposium, {Abundances in the Galactic Center}.
pp 107--119

\bibitem[\protect\citeauthoryear{{Yusef-Zadeh}, {Braatz}, {Wardle} \&
  {Roberts}}{{Yusef-Zadeh} et~al.}{2008}]{FYZ2008}
{Yusef-Zadeh} F.,  {Braatz} J.,  {Wardle} M.,    {Roberts} D.,  2008, \apj,
  683, L147

\bibitem[\protect\citeauthoryear{{Yusef-Zadeh}, {Roberts}, {Goss}, {Frail} \&
  {Green}}{{Yusef-Zadeh} et~al.}{1999}]{FYZ1999}
{Yusef-Zadeh} F.,  {Roberts} D.~A.,  {Goss} W.~M.,  {Frail} D.~A.,    {Green}
  A.~J.,  1999, \apj, 512, 230

\bibitem[\protect\citeauthoryear{{Yusef-Zadeh}, {Stolovy}, {Burton}, {Wardle}
  \& {Ashley}}{{Yusef-Zadeh} et~al.}{2001}]{FYZ2001}
{Yusef-Zadeh} F.,  {Stolovy} S.~R.,  {Burton} M.,  {Wardle} M.,    {Ashley}
  M.~C.~B.,  2001, \apj, 560, 749

\bibitem[\protect\citeauthoryear{{Zhao}, {Morris}, {Goss} \& {An}}{{Zhao}
  et~al.}{2009}]{Zhao2009}
{Zhao} J.,  {Morris} M.~R.,  {Goss} W.~M.,    {An} T.,  2009, \apj, 699, 186

\end{thebibliography}
\clearpage
\newpage

\appendix 

\section{Disk Parameters}
\subsection{Orientation}
\paragraph*{}
The direction of the disk's major axis is usually specified by its PA East of
North, which in this study equates to the complementary angle of $\phi$, and its
inclination to the plane of the sky, $\alpha$, as illustrated in Fig.\ \ref{CND
params}.  The disk's axes are designated as x$_{D}$ with the positive to the
left of centre, y$_{D}$ positive to the top of the figure and z$_{D}$ positive
away form an observer observer's view along the axis of rotation. Common values of the parameters are given in Table \ref{Ring pars}. 
\paragraph*{}

\begin{figure} 
%\vspace{40mm}
\includegraphics[bb= -2 275 600 670,scale=0.4]{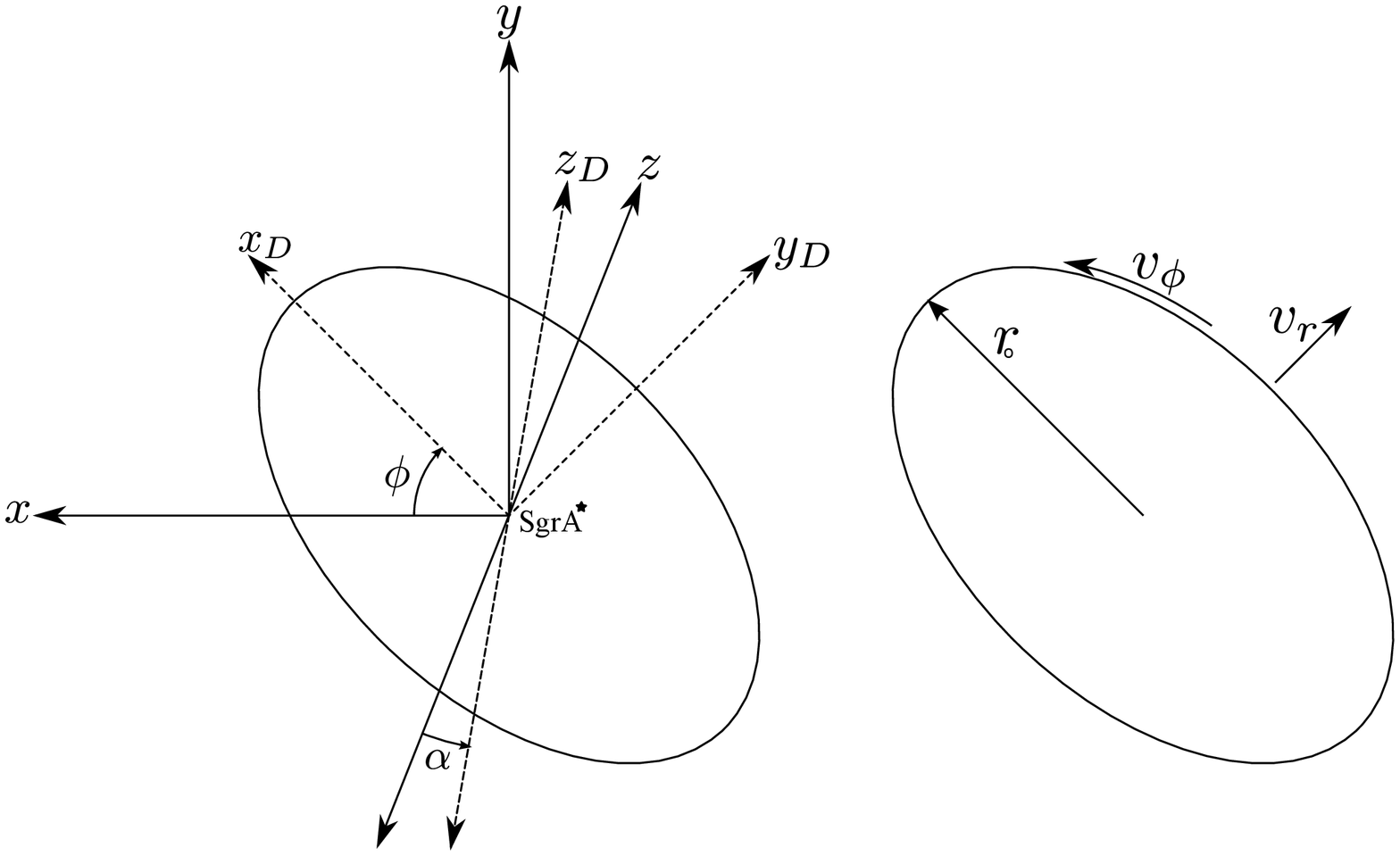} 
\centering
\fontsize{9} {9} 
\caption[Illustration of CND's defining parameters]{\label{CND params}Illustrates the CND's defining parameters, the left diagram shows the plane of the sky xy with the z axis as the line of sight. x$_{D}$, y$_{D}$ define the disk's plane with z$_{D}$ the axis of rotation, $\alpha$ is the angle between z and z$ _{D} $; $\phi$ is the angle of the disk's major axis measured clockwise from East in the plane of the sky xy and is the complement of the PA (which is measured East of North). 
The right hand diagram shows the disk's physical dimensions: r$_{0} $ is the mean radius of the ring; v$ _{\phi} $ is the rotational velocity; v$ _{r} $ is los velocity, $\Delta$r is radial thickness with FWHM thickness $ = r_{0}\sqrt{\mathrm{log4}} = 1.18r_{0} $. The thickness perpendicular to the plane of the ring and is assumed to be zero. Typical values for these parameters are listed in Table \ref{Ring pars}. This figure and Table \ref{Ring pars} are based on Fig.\ 9 and Tables 1 \& 2 of \citet{Marshall1995}.} 

\end{figure}

\begin{table*} 
\begin{minipage}{140mm}
\centering 
\caption{\label{Ring pars} CND's Defining Parameters.}
\begin{tabular}{ccccccc} 
\hline
\multicolumn{1}{c}{ }&\multicolumn{1}{c}{$\alpha$}&\multicolumn{1}{c}{$\phi$ }&\multicolumn{1}{c}{v$_{\phi}$}&\multicolumn{1}{c}{v$_{r}$}&\multicolumn{1}{c}{r$_{0}$}&\multicolumn{1}{c}{v$_{disp}$} \\
 &\multicolumn{1}{c}{(deg)}&\multicolumn{1}{c}{(deg)}&\multicolumn{1}{c}{(km s$^{-1}$)}&\multicolumn{1}{c}{(km s$^{-1}$)} &\multicolumn{1}{c}{(km s$^{-1}$)} \\
  & & & & & & \\
\hline
\multicolumn{1}{c}{Previous Estimates$^{1}$}&60-70&60-65&100-110&{$<20$}&40-50&10-70 \\
{Larger Ring$^{2}$}&70& &110&52&96&10 \\
\multicolumn{1}{c}{HCN$_{3-2}$ $^{3}$}&59&64&76&-12&38&47 \\ 
\multicolumn{1}{c}{HCN$_{4-3}$ $^{3}$}&60&65&72&-13&44&40 \\ 
\multicolumn{1}{c}{HCN$_{3-2}$ $^{4}$}&70$ \pm 5 $&$ \sim65 $&110$ \pm 5 $&$ < 19$&39-52& \\
\hline

\multicolumn{7}{c}{[1] based on \citet{Harris1985,Guesten1987,Sutton1990,Jacks1993}} \\
\multicolumn{7}{l}{[2] HI absorption,\citet{Liszt1985}} \\
\multicolumn{7}{l}{[3] \citet{Marshall1995}} \\
\multicolumn{7}{l}{[4] \citet{Jacks1993}} \\
\end{tabular}
\end{minipage}
\end{table*} 

\subsection{Co-ordinate Transformation} \label{cordtran}
\paragraph*{}
Astronomical figures use a 3D axes convention where the x is positive in the easterly direction, the y axis is positive in the northerly direction and the line of sight (z) axis, is positive in the direction away from the observer. The disk co-ordinates have the subscript,D, to distinguish them from the sky co-ordinates.

\paragraph*{}
Rotation matrices can be used to transform plane of sky to plane of disk co-ordinates where co-ordinates in this paper are expressed as offsets from SgrA*. The process is performed in two rotations, the first about the the line of sight (oz) so that the new axes, x$^{'}$ and  y$^{'}$ are at an angle    
$\phi$  to the xy axes and align with the major and minor axes of the projected disk and the second about the ox$^{'}$ axis by an angle $\alpha$ to align a viewer's line of sight with the disk's axis of rotation to produce a projection of the disk with true or deprojected dimensions and  clump offsets, x$_{D}$ and y$_{D}$ from SgrA$^{*}$.

The first transformation is given by
\begin{equation} \label{xytran}
\left( \begin{array}{c}x'\\y'\\z' \end{array}\right) = \left( \begin{array}{ccc}\cos\phi &\sin\phi &0\\-\sin\phi &\cos\phi &0\\0&0&1
\end{array} \right)\left( \begin{array}{c}x\\y\\z \end{array}\right) \,.
\end{equation}

This second transformation is given by

\begin{equation} \label{yztran}
\left( \begin{array}{c}x_{D}\\y_{D}\\z_{D} \end{array}\right) = \left( \begin{array}{ccc}1&0&0\\0&\cos\alpha&-\sin\alpha\\0&\sin \alpha&\cos\alpha \end{array} \right) \left( \begin{array}{c}x'\\y'\\z' \end{array}\right) \,.
\end{equation}
\paragraph*{}
The transformation from sky to disk co-ordinates is then derived by substituting the matrices from Eqn.\ \ref{xytran} into Eqn.\ \ref{yztran} to produce

\begin{equation}\label{TM}
 \left( \begin{array}{c}x_{D}\\y_{D}\\z_{D} \end{array}\right)=\left(
\begin{array}{ccc} \cos\phi \sin\phi&\sin\phi&0\\-\sin\phi\cos\alpha&\cos\phi\cos\alpha&-\sin\alpha
\\-\sin\phi\sin\alpha& \cos\phi\sin\alpha&\cos\alpha \end{array}\right)
\left( \begin{array}{c}x\\y\\z \end{array}\right) \,. \\
\end{equation}

The inverse transformation of Eqn \ref{TM} is

\begin{equation} \label{ITM}
\left( \begin{array}{c}x\\y\\z \end{array}\right)=\left(
\begin{array}{ccc} \cos\phi &-\sin\phi\cos\alpha&-\sin\phi\sin\alpha\\\sin\phi&\cos\phi\cos\alpha&\cos\phi\sin\alpha
\\0& -\sin\alpha&\cos\alpha \end{array}\right)
\left( \begin{array}{c}x_{D}\\y_{D}\\z_{D} \end{array}\right) \,, \\
\end{equation}

where x$_{D}$ and y$_{D}$, x and y offset  are offsets from SgrA$^{*}$ and z$_{D}$ is the disk's z co-ordinate which is aligned with its axis of rotation.

\subsection{Line of Sight Velocity} \label{radv}

\paragraph*{}
A clump's line of sight velocity along the line of sight velocity is caused by the change in position of the clump in the disk's plane generating a change in the distance along the line of sight in the sky. The position of a clump in the disk is specified by its x$_{D}$ and y$_{D}$ co-ordinates. Evaluation of x$_{D}$ is straightforward as it is only dependent on the projected x and y co-ordinates. The y$_{D}$ co-ordinate is dependent on x, y and z co-ordinate values (see Eqn.\ \ref{ITM}). The projected z co-ordinate along the line of sight is not directly observable, but can be expressed as a function of the projected x and y co-ordinate values and y$_{D}$ calculated by assuming z$_{D}$ is zero as there is no independent information available for this quantity and it seems reasonable to adopt this assumption. 

\paragraph*{}  
The ''line of sight`` velocity from the flat rotation model is then obtained from the following expression:
\begin{eqnarray}\label{RadV2}
 v_z = \frac{\mathrm{d}z}{\mathrm{d}t} = \mathrm{v}_\phi\cos \phi_D\sin \alpha \,.
\end{eqnarray}

\normalsize
\bsp

\label{lastpage}

\end{document}